\documentclass[dvipsnames,letterpaper, 10 pt, journal]{ieeetran}

\IEEEoverridecommandlockouts	
\usepackage{algorithm}
\usepackage{algpseudocode}
\usepackage{cite}
\usepackage{amsmath,amssymb,amsfonts,amsthm}
\usepackage{textcomp}
\usepackage{graphicx,color}
\usepackage{mathrsfs}
\usepackage{tikz}
\usepackage{subfigure}
\usepackage{url}
\usepackage{color}
\usepackage{dsfont}
\usepackage{bbm}
\usepackage{booktabs}
\usepackage{array}
\usepackage{yfonts}
\usepackage[normalem]{ulem}
\usepackage{arydshln,leftidx,mathtools}
\usepackage{multicol}
\usepackage{amsmath}
\usepackage{stfloats}
\usepackage{comment}
\usepackage{subfigure}
\usepackage{manyfoot}
\usepackage{enumitem}

\definecolor{myBlue}{RGB}{49,130,189}

\newtheorem{theorem}{Theorem}[section]

\newtheorem{remark}{Remark}
\newtheorem{assumption}[theorem]{Assumption}

\newcommand{\map}[3]{#1: #2 \rightarrow #3}

\newcommand{\blkdiag}{\mathrm{blkdiag}}

\usepackage{tabularx}
\usepackage{multirow}
\usepackage{makecell}

\newcommand\aamsout{\bgroup\markoverwith{\textcolor{violet}{\rule[0.5ex]{2pt}{1pt}}}\ULon}

\newcommand{\real}{\mathbb{R}}

\newcommand{\T}{\mathsf{T}} 

\newcommand{\mc}{\mathcal}

\newcommand{\expect}[1]{\mathbb{E}\left[#1\right]}

\newcommand{\1}{\mathds{1} }

\DeclareSymbolFont{bbold}{U}{bbold}{m}{n}
\DeclareSymbolFontAlphabet{\mathbbold}{bbold}

\newcommand\oprocendsymbol{\hbox{$\square$}}
\newcommand\oprocend{\relax\ifmmode\else\unskip\hfill\fi\oprocendsymbol}

\renewcommand{\theenumi}{(\roman{enumi})}

\newcommand*{\QEDA}{\hfill\ensuremath{\blacksquare}}%

\DeclareNewFootnote{R}[roman]

\graphicspath{{figs/}}

\makeatletter
\let\NAT@parse\undefined
\makeatother
\usepackage[colorlinks,urlcolor=blue, linkcolor=blue, citecolor=blue]{hyperref}

\renewcommand{\baselinestretch}{0.9841}

\begin{document}

\title{\LARGE \bf DER Allocation without Load Prediction via Reinforcement Learning}

\author{Abed~AlRahman~Al~Makdah, Aravind~Ramana, Shaofeng~Zou, Oliver~Kosut, and Lalitha~Sankar
  \thanks{This material is based upon work supported by the National Science Foundation under Grant No. EPCN-2246658 and a PSERC grant S-114. A. A. Al Makdah, S. Zou, O. Kosut, and L. Sankar are with the School of Electrical, Computer and Energy Engineering at the Arizona State University,
    \href{mailto:aalmakda@asu.edu}{\{\texttt{aalmakda}},\href{mailto:zou@asu.edu}{\texttt{zou}},\href{mailto:okosut@asu.edu}{\texttt{okosut}},\href{mailto:lsankar@asu.edu}{\texttt{lsankar\}@asu.edu}}. A. Ramana is with the Department of Physics at the Indian Institute of Technology Madras, \href{mailto:ep23b003@smail.iitm.ac.in}{\texttt{ep23b003@smail.iitm.ac.in}}.}}

\maketitle
\pagestyle{empty}
\thispagestyle{empty}

\begin{abstract}
The growing variability of renewable generation increases the need for fast and flexible grid-balancing mechanisms. Existing frameworks for distributed energy resource aggregations (DERAs) rely on short-term forecasts of net demand, making their performance highly sensitive to prediction errors. In this paper we present a forecast-free reinforcement learning (RL) framework for DERA allocation that learns optimal policies directly from operational data. We model the DERA dynamics as a deterministic linear system and the exogenous net load as a feature-based linear Markov process, capturing short-range temporal dependencies without explicit forecasting. We derive a closed-form expression for the optimal policy, which is learned through a least-squares value iteration (LSVI) algorithm using data collected across episodes. The proposed framework preserves the interpretability and constraint satisfaction of DER model while adapting to stochastic demand variations through data-driven updates. Numerical experiments on real California Independent System Operator (CAISO) net-demand data demonstrate that the learned controller achieves high tracking accuracy and stable regulation across heterogeneous DER aggregators without requiring any demand prediction.
\end{abstract}


\section{Introduction}\label{sec: introduction}
The growth of renewable generation introduces uncertainty in the power grid leading to increased variability and ramping requirements in net demand. Traditionally, such variability is addressed using fossil fuel-based generators such as coal and natural gas, which are both costly and environmentally unsustainable. A promising alternative is to leverage distributed energy resources (DERs) --- including buildings, thermostatically controlled loads, storage devices, and electric vehicles (EV)---via DER aggregators (DERAs). An aggregator can provide fast, flexible responses when it coordinates across different DER types effectively. Current approaches to coordinating DERs are predominantly forecast-driven. Forecast errors not only reduce tracking performance but also complicate aggregator operation, as errors propagate through ramping and state-of-charge dynamics.

In this paper, we develop a forecast-free alternative. In particular, we develop a reinforcement learning (RL) framework that learns an allocation policy directly from operational data, rather than predicting future demand. We model the net-load as a feature-based linear Markov process that captures its short-range dependence. At a DERA, each DER is represented by a generalized battery model with state-of-charge leakage and ramp-rate dynamics, yielding a tractable deterministic linear system. Combining the two yields a hybrid structure involving the classical deterministic linear dynamics of the device combined with a stochastic Markovian demand. The resulting model allows us to combine a quadratic tracking objective that preserves the analytical simplicity of the linear quadratic regulator (LQR) while exploiting policy updates using reinforcement learning.

\textbf{Related work.}
A substantial portion of the literature relies on forecast-based, finite-horizon optimal control. In \cite{EB-MC-ED:19}, a nonlinear AC optimal power flow including thermostatically controlled loads is cast as a Markov decision process and reformulated as a convex finite-horizon problem. Similar setups appear in \cite{NC-JM-MK-AB-AM:18,JM-SM-HB-MA:22}. In \cite{JM-RA-OK-LS:24} , a centralized model predictive control (MPC) framework is used for DERA allocation. Distributed MPC formulations have been proposed for economic dispatch \cite{JM-MAM-NL-FA:17} and frequency regulation \cite{ML-YS-XL:16, MM-CZ-XL-HC:17, AP-CW-TK-JE-KS-KHJ:17}, but these works either omit DERs or focus on device-level tracking without aggregate battery models or grid-level allocation. In \cite{LADE-MA:20}, a packetized energy management (PEM) Macro model approach is introduced. Unlike forecast-based or finite-horizon optimization schemes, the PEM framework matches demand using myopic, real-time modulation of packet acceptance probabilities rather than solving predictive control problems. Recent work has increasingly explored reinforcement learning and data-driven methods for grid flexibility and distributed energy resource coordination. In \cite{DC-JZ-WH-FD-QH-ZC-FB:21}, a multi-agent deep RL strategy for coordinating inverter-based DERs to maintain voltage stability under high renewable penetration is proposed, which demonstrates the ability of RL to handle nonlinearity and uncertainty in distribution networks. In \cite{SB-YCC-VWSW:21}, a deep RL–based residential demand response controller that learns consumption strategies under stochastic load and price signals is developed. In \cite{HL-JQ-JZ:22}, a data-driven scheduling model for a virtual power plant that participates in day-ahead and real-time markets is proposed. These formulations generally rely on explicit forecasts or stochastic models of prices, loads, or renewable outputs and often optimize economic or network-regulation objectives over long horizons. In contrast, our work develops a forecast-free RL controller that directly allocates DERA in real time to track an exogenous load demand signal, without relying on load forecasting. This shifts the focus to model-free demand matching using online data.

\textbf{Contributions.} This paper features three contributions. First, we eliminate the need for demand forecasts by coupling DERA dynamics with a feature-based linear Markov model for the net-load, allowing adaptive allocation policies. Second, we derive a closed-form expression for the optimal policy, which is learned and updated using online data, without requiring explicit identification of the demand model. Finally, we demonstrate the effectiveness of our framework through numerical experiments on real net-load data.

\textbf{Notation.} A Gaussian random variable $x$ with mean $\mu$ and
covariance $\Sigma$ is denoted as $x\sim\mc{N}(\mu,\Sigma)$. The
$n\times n$ identity matrix is denoted by $I_n$. The expectation operator is
denoted by $\mathbb{E}[\cdot]$. A positive definite (semidefinite) matrix $A$ is
denoted as $A\succ 0$ ($A\succeq 0$). The Kronecker product is denoted
by $\otimes$.

\section{Problem formulation}\label{sec: formulation}
 We consider a generalized battery model for each aggregator. The $i$-th DER aggregator obeys:
\begin{align}\label{eq: DER_model}
\begin{split}
z_{i,t+1}   &= \alpha_i z_{i,t}   - \beta_i p_{i,t},  \quad \!\! t\!\in\!\{0,\ldots,T\!-\!1\},\\
p_{i,t+1}&= p_{i,t} + u_{i,t}
\end{split}
\end{align}
where $z_{i,t}\in \mathbb{R}$ is the state of charge (SoC), $\beta_i \in \mathbb{R}$ is a sampling time constant, $\alpha_i\in [0,1]$ is a leakage parameter, $p_{i,t} \in \mathbb{R}$ is the power supplied, and $u_{i,t} \in \mathbb{R}$ is the ramp rate at time $t\geq 0$. We re-write \eqref{eq: DER_model} in linear state-space form~as
\begin{align*}
\underbrace{\begin{bmatrix}
 z_{i,t+1}\\
 p_{i,t+1}
\end{bmatrix}}_{x_{i,t+1}}
=
\underbrace{\begin{bmatrix}
 \alpha_i & -\beta_i\\
 0 & 1
\end{bmatrix}}_{A_i}
\begin{bmatrix}
 z_{i,t}\\
 p_{i,t}
\end{bmatrix}
+
\underbrace{\begin{bmatrix}
 0 \\
 1
\end{bmatrix}}_{B_i}
u_{i,t}.
\end{align*}
For $M$ aggregators, the complete linear state space model is
\begin{align}\label{eq: system}
\begin{split}
\underbrace{\begin{bmatrix}
 x_{1,t+1}\\
\vdots\\
 x_{M,t+1}
\end{bmatrix}}_{x_{t+1}}
=&
\underbrace{\begin{bmatrix}
 A_1 &\cdots & 0\\
 \vdots & \ddots  &\vdots\\
 0 & \cdots & A_M
\end{bmatrix}}_{A}
\begin{bmatrix}
 x_{1,t}\\
 \vdots \\
 x_{M,t}
\end{bmatrix}\\
&+
\underbrace{\begin{bmatrix}
 B_1 &\cdots & 0\\
 \vdots & \ddots  &\vdots\\
 0 & \cdots & B_M
\end{bmatrix}}_{B}
\begin{bmatrix}
u_{1,t}\\
\vdots\\
u_{M,t} 
\end{bmatrix},
\end{split}
\end{align}
where $x_t\in \mathcal{X} =\mathbb{R}^{2M}$ and $u_t \in \mathcal{U} =\mathbb{R}^M$. Let 
\begin{align}\label{eq: system_output}
\begin{split}
y_t=&\underbrace{\Big({\1_M}^{\T} \otimes
\begin{bmatrix}
 0 &1
\end{bmatrix}\Big)}_{C} x_t,
\end{split}
\end{align}
where $y_t\in \mathbb{R}$ is the total power supplied by the aggregators. We consider net load demand evolving according to the discrete-time Markov~process
\begin{align}\label{eq: load}
 s_{t+1}| s_t\! \sim\! \mathbb{P}_t\left(s_{t+1}| s_t\right), \quad \!\! t\!\in\!\{0,\ldots,T\!-\!1\},
\end{align}
where $s\in \mathcal{S} \subseteq \real$ denotes net load demand and $\mathbb{P}_t\left(s'|s\right)$ denotes the transition probability from state $s$ to $s'$.  We introduce the following assumption on the Markov Process~in~\eqref{eq: load}.
\begin{assumption}{\bf \emph{(Linear Markov Process)}}\label{assump: linMDP}
Let $w\in \mathcal{S}^r$, $\map{\phi}{\!\mathcal{S}^r}{\!\real^d}$ be a known feature vector, and $\mu_t\!\in\! \mathbb{R}^d$ a vector of $d$ unknown signed measures over $\mathcal{S}$. For $s\!\in\! \mathcal{S}$, we have
\begin{align}
\begin{bmatrix}
 s_{t-r+2}\\
 \vdots \\
 s_t\\
 s_{t+1}
\end{bmatrix}
=
\underbrace{\begin{bmatrix}
 0 &1 & \cdots & 0\\
 \vdots & \ddots & \ddots & \vdots\\
 0 & 0 & \cdots & 1\\
  0 & 0 & \cdots & 0
\end{bmatrix}}_{\widehat{A}}
\underbrace{\begin{bmatrix}
 s_{t-r+1}\\
 \vdots \\
 s_{t-1}\\
 s_{t}
\end{bmatrix}}_{w_t}
+
\underbrace{\begin{bmatrix}
 0\\
 \vdots \\
 0\\
1
\end{bmatrix}}_{\widehat{B}} s_{t+1},
\end{align}
\begin{align}\label{eq: linApprox}
 \mathbb{P}_t\left(s_{t+1}|w_t\right)=\phi(w_t)^{\T} \mu_t(s_{t+1}),
\end{align}
We assume $\|\phi(w)\|\leq 1/\sqrt{d}$ and $\|s\|\leq \delta_s$ for all $s \in \mc{S}$, $\expect{\phi(w_t)\phi(w_t)^{\T}}\succ 0$, and $\|\mu_t\|\leq 1$ for all $t$.
\end{assumption}
We consider the cost following cost at time $t$,
\begin{align}\label{eq: cost}
 c(x_t,s_t,u_t)=&{x_t}^{\T}W_x{x_t} + u_t^{\T}Ru_t + \kappa \left(s_t-y_t\right)^2
\end{align}
where $W_x\succeq 0$, $R\succ 0$, and $\kappa >0$. The cost in \eqref{eq: cost} can be written in the following form
\begin{align*}
 c(x_t, s_t, u_t) =&
 \begin{bmatrix}
 x_t \\  s_t \\ u_t
\end{bmatrix}^{\T}
\underbrace{\begin{bmatrix}
W & F & D \\
 F^{\T} & M & H \\
 D^{\T}    & H^{\T}    & R \\
\end{bmatrix}}_{P}
 \begin{bmatrix}
 x_t \\  s_t \\ u_t
\end{bmatrix},
\end{align*}
with $W=W_x+\kappa C^{\T} C$, $M=\kappa$, $F=-\kappa C^{\T}$, $D=0$, and $H=0$.
The aggregators follow a control policy $\map{\pi_t}{\mathcal{X}\times \mathcal{S}^r}{\mathcal{U}}$, where $u_t=\pi_t\left(x_t,w_t\right)$ is the action that the agent takes at state $x_t$ and $s_t$ at time $t$, for $t\geq0$. We seek to find an optimal control policy, $\boldsymbol{\pi}=(\pi_{0},\ldots,\pi_{T})$ for the following task:
\begin{align}\label{eq: control task}
\begin{array}{ll}
   	\underset{\boldsymbol{\pi}}{\text{minimize}} & \displaystyle \expect{\sum_{t=0}^{T} c\left(x_t,s_t,u_t\right)}, \\
   	\text{subject to} & 
	x_{t+1} = A x_t+B u_t,\\
	& s_{t+1} \sim \mathbb{P}_t\left(s_{t+1}|w_t\right), \\
	& w_{t+1}=\widehat{A} w_t +\widehat{B} s_{t+1}\\
	 &u_t = \pi_t\left(x_t, w_t\right),
	  \end{array}
\end{align}
with $u_T=\pi_T\left(x_T,w_T\right)=0$. 
We define the state-action value function $\map{Q_{t}^{\pi}}{\mathcal{X} \times  \mathcal{S}^r \times \mathcal{U}}{\real}$ as the expected cumulative cost under policy $\pi$ starting from state $x_t$, $s_t$, $w_t$, and action $u_t$ at time $t$, given~by
\begin{align*}
&Q_{t}^{\pi}\!\left(x,s,w,u\right)\triangleq c\! \left(x,s, u \right)\\
& + \expect{\sum_{i=t+1}^{T}\! \!c\! \left(x_{i},s_{i}, \pi_t\left(x_{i},w_{i}\right)\right) \!\Big| x_t\!=\!x, s_t\!=\!s, w_t\!=\!w , u_t\!=\!u}\!\!.
\end{align*}

\section{Reinforcement Learning Algorithm}\label{sec: main results}

In this section, we leverage the hybrid structure, deterministic linear DERA dynamics coupled with a feature-based linear Markov model of the net load and the quadratic structure of the cost, to derive a parametric expression for the optimal policy. Then, we introduce a least-squares value iteration (LSVI) algorithm to learn the parameters from online data.
In \cite{AAAM-OK-LS-SZ:25}, we derived a parametric expression of the state-action value function that is linear in the feature map, $\phi$, along with a parametric expression for the corresponding optimal policy. In the next result, we present the parametrized expression of optimal policy, which is adopted from \cite{AAAM-OK-LS-SZ:25}.

\begin{theorem}{\bf \emph{(Optimal policy representation \cite{AAAM-OK-LS-SZ:25})}}\label{thrm: opt_policy}
For any $x\in \mc{X}$, $s\in \mc{S}$, $w\in \mc{S}^r$, and~$t \in \{0,\cdots, T-1\}$
\begin{align}\label{eq: optimal policy}
\begin{split}
u_t^*(x,w)=&\pi_t^*(x,w)\\
=&K_{t,x}x + K_{t,s}s + K_{t,h} \left({\phi(w)}^{\T} \otimes Z\right) \theta_{t+1},
\end{split}
\end{align}
where $Z\!=\![I_n, 0_{n \times 1}]$,
\begin{align}\label{eq: feedback gains}
\begin{split}
 K_{t,x}&=-\left(R +B^{\T} G_{t+1} B \right)^{-1} \left( B^{\T}G_{t+1}A+D^{\T}\right),\\
 K_{t,s}&=-\left(R +B^{\T} G_{t+1} B \right)^{-1}H^{\T},\\
 K_{t,h}&=-\left(R +B^{\T} G_{t+1} B \right)^{-1}B^{\T},
 \end{split}
\end{align}
 and $G_{t+1}$ satisfies,
\begin{align}\label{eq: riccati}
 G_t=&A^{\T}G_{t+1}A + W \\
 - (A^{\T}&G_{t+1}B \!+\! D) (R\!+\!B^{\T}G_{t+1}B )^{-1} (B^{\T}G_{t+1}A \!+\! D^{\T}),\nonumber
\end{align}
and $\theta_{t+1} \in \mathbb{R}^{d(n+1)}$.
\end{theorem}
Theorem \ref{thrm: opt_policy} shows that the optimal control at time $t$ admits a feedback form in the physical DERA state, $x$, the current net load, $s$, and a feature summary of recent net-load history, $w$. The feedback gains, $K_{t,x}$ and $K_{t,s}$ are LQR-type expressions computed from a Riccati recursion for  that incorporates the tracking cross-terms, while the feature feedforward term, $K_{t,h}$ captures how the linear Markov process shapes future costs through parameters $\theta_{t+1}$. Next, we leverage this structure and propose a least-squares value iteration algorithm to learn $\theta$ directly from data, avoiding explicit demand forecasting.
\begin{algorithm}[b]
    \caption{Least-Squares Value Iteration}
    \label{alg: alg1}
    \begin{algorithmic}[1]
     \State Given: $L$, $R_{\theta}$, $\lambda$
            \For{episode $\ell=1,\cdots , L$\hspace{5pt}}\\
		\hspace{12pt} $x^{\ell}_0 = x^{\ell-1}_T$\\
		\hspace{12pt} $w^{\ell}_0 = w^{\ell-1}_T$
                \For{step $t=T-1,\cdots , 0$\hspace{5pt}}
                    \State {\small{$\Lambda_t^{\ell}\!\gets\! {\!\sum_{i\!=\!1}^{\ell\!-\!1}\!\!{Y\!(x^{i}_t,\!u^{i}_t)}^{\T}\! \phi(w^{i}_t) {\phi(w^{i}_t)}^{\T}\! Y\!(x^{i}_t,\!u^{i}_t) \! +\! \lambda I_{dn+d}}$}}
                    \State {\scriptsize{$\theta_{t+1}^{\ell}\! \gets\! (\Lambda_t^{\ell})^{-1} \!\!\sum_{i\!=\!1}^{\ell\!-\!1}\!{Y\!(x^{i}_t,\!u^{i}_t)}^{\T}\! \phi(w^{i}_t)\epsilon_{t+1}^{\ell}\!(x^{i}_{t+1},\!s^{i}_{t+1},\!w^{i}_{t+1})$}}
                    \If{$\|\theta_{t+1}^{\ell}\|>R_{\theta}$}
			\State $\theta_{t+1}^{\ell} \gets \frac{R_{\theta}}{\|\theta_{t+1}^{\ell}\|}\theta_{t+1}^{\ell}$
		   \EndIf
                \EndFor
                
                \For{step $t=0,\cdots , T-1$\hspace{5pt}}
                \State $u^{\ell}_t \gets K_{t,x} x^{\ell}_t + K_{t,s} s^{\ell}_t + K_{t,h} ({\phi(w^{\ell}_t)}^{\T} \otimes Z )\theta_{t+1}^{\ell}$
                \State Take action $u^{\ell}_t$
                \State Observe $x^{\ell}_{t+1}$, $s^{\ell}_{t+1}$, and $w^{\ell}_{t+1}$
                \EndFor
            \EndFor
    \end{algorithmic}
\end{algorithm}
We adapt the least-squares value iteration algorithm (Algorithm \ref{alg: alg1}) from our prior work \cite{AAAM-OK-LS-SZ:25}. Our algorithm consists~of an outer loop over $L$ episodes, where each episode consists of two loops: 1) backward-in-time weight update loop (lines 5-11) and 2) forward roll-out and data collection loop (lines 12-16). During the first pass of episode~$\ell$ (lines 5–11), we treat the data collected in the previous $\ell\!-\!1$ episodes as a fixed dataset
\begin{align}\label{eq: dataset}
 \mathcal D_{\ell-1}\!:=\!\bigl\{
        (x^{i}_t,s^{i}_t,u^{i}_t,x^{i}_{t\!+\!1},s^{i}_{t\!+\!1})
        :i<\ell,\;0\leq t<T
     \bigr\}.
\end{align} 
At each time step $t$, $\theta$ minimizes a regularized least-squares loss. Solving this problem on past trajectory data yields an accurate value-function approximation and enables closed-form greedy policy updates without estimating the transition probabilities. The regularized least-squares regression is given by the following optimization solved in Algorithm \ref{alg: alg1}.
 \begin{align}
  \theta^{\ell}_{t+1}\!\!=\!&
      \underset{\theta\in\mathbb R^{d(n+1)}}{\arg\min}
      \sum_{i=1}^{\ell-1}
            \!\Bigl(
                \phi(w_t^{i})^{\T} Y(x_t^{i},u_t^{i}) \theta \\
                &\qquad\qquad  \qquad-\!
               \epsilon_{t+1}^{\ell}(x^i_{t+1},s^i_{t+1},w^i_{t+1})
            \Bigr)^{2}
      +\lambda\|\theta\|^{2},\nonumber
\end{align}
where
\begin{align}
\epsilon_{t+1}^{\ell}\left(x,s,w\right) &= 2\left(x\right)^{\T}h^{\ell}_{t+1}\left(s,w\right) + q^{\ell}_{t+1}\left(s,w\right),\\
h^{\ell}_{t+1}\left(s,w\right)&=\left(A^{\T} + K_{t,x}^{\T}B^{\T}\right)(\phi(w)^{\T} \otimes Z) \theta^{\ell}_{t+2}\\
&\quad+ \left(F +K_{t,x}^{\T}H^{\T}\right)s,\nonumber\\
q^{\ell}_{t+1}(s,w)&=\left({\phi(w)}^{\T} \otimes \overline{Z}\right) \theta^{\ell}_{t+2} +s^{\T}(M + H K_{t,s}) s\nonumber\\
 &\quad+ {\theta^{\ell}_{t+2}}^{\T} \left({\phi(w)}\!\otimes\! Z^{\T}\right) B K_{t,h} ({\phi(w)}^{\T} \!\otimes\! Z) \theta^{\ell}_{t+2}\nonumber\\
 &\quad+2 s^{\T}H K_{t,h}\left({\phi(w)}^{\T} \otimes Z\right) \theta^{\ell}_{t+2},
\end{align}
with $Y(x,u)\!=\!I_d\!\otimes\! [2 \left(Ax + Bu\right)^{\T} , 1]$, $Z\!=\![I_n, 0_{n \times 1}]$, and $\overline{Z}=[0_{1\times n}, 1]$. And $\lambda>0$ is the regularization parameter. The closed-form parameter update are given by lines 6 and 7 of Algorithm \ref{alg: alg1}. At episode $\ell=1$, we assume that we have access to the net-load values from time $t=-r+1$ to $t=0$ so that the initial vector $w_0^{\ell}$ is well defined for any $r\geq 1$. Further, at $\ell=1$, we set $x_0^{\ell}\sim \mathcal{N}(0,I_n)$, $\theta^{\ell}_{t+1}\sim \mathcal{N}(0,I_{dn+d})$ and $\Lambda_t^{\ell}=\lambda I_{d(n+1)}$ for $t\in \{0,\cdots, T-1\}$. The regularizer term $\lambda I_{d(n+1)}$ ensures numerical stability, the projection step in lines 8-10 makes sure that the norm of the learned parameters is uniformly bounded for $t\in \{0,\cdots,T-1\}$ and $\ell \in\{1,\cdots ,L\}$. In the second pass (lines 12–16) the newly computed parameters $\theta^{\ell}_{t+1}$ are plugged into the greedy closed-form policy \eqref{eq: optimal policy} as shown in step 13 of Algorithm \ref{alg: alg1} to
generate a new trajectory $(\{x^{\ell}_t,s^{\ell}_t,u^{\ell}_t\}_{t=0}^{T})$. These samples are appended to the collected data \eqref{eq: dataset}, and will be used in the next episode’s backward update. 

The control formulation in this paper builds on our framework in \cite{AAAM-OK-LS-SZ:25}, which establishes closed-form optimal policies for linear dynamics driven by feature-based Markov processes. In this work, we specialize our framework in \cite{AAAM-OK-LS-SZ:25} to real-time grid balancing with heterogeneous DER aggregations. In particular, we introduce a generalized battery and ramping model for DERAs, a forecast-free net-load tracking formulation using real CAISO data, and a regulation-oriented objective aligned with system operator dispatch. This enables real-time DERA allocation without short-term demand forecasting while preserving the structure and interpretability of classical LQR controllers. Furthermore, unlike \cite{AAAM-OK-LS-SZ:25}, in this paper we model the evolution of the exogenous state $s_{t+1}$ to be conditioned on the window $w_t$, which captures the short-range temporal dependencies required by the linear Markov process.
\section{Numerical experiments}\label{sec: numerical experiments}
We adopt a similar experimental setup to \cite{JM-RA-OK-LS:24}. For reference, we compare against a forecast-based MPC benchmark, similar to \cite{JM-RA-OK-LS:24}, which computes DERA allocations using short-term net-load predictions. Implementation details are omitted since MPC is used only as a baseline. We evaluate our forecast-free reinforcement learning framework using real net-demand data from the California Independent System Operator (CAISO) for the period from May 15, 2023 to September 15, 2023 \cite{CAISO}. The dataset contains net-load measurements (total demand minus renewable generation) sampled at each $5$ minutes, represented by the matrix $S\in \mathbb{R}^{1\times N_d}$, where $N_d$ denotes the total number of samples, which is written as
\begin{align}\label{eq: net load data}
 S=
\begin{bmatrix}
 s_{-r+1} &\cdots&s_0& \cdots & s_{N_d}
\end{bmatrix}.
\end{align}
\begin{figure}[!t]
  \centering
  \includegraphics[width=0.95\columnwidth,trim={0cm 0cm 0cm
    0cm},clip]{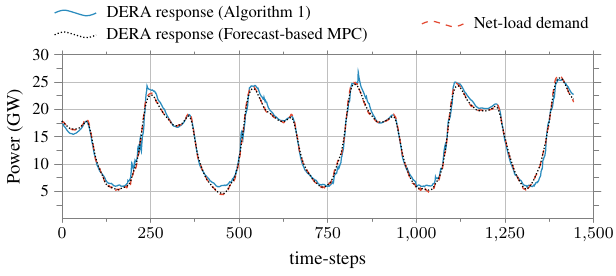}
  \caption{This figure shows the net-load demand (dashed red line), the cumulative DERA response under the proposed forecast-free RL controller (solid blue line), and the cumulative DERA response under a forecast-based MPC benchmark (dotted black line) over a five day period, where each time-step corresponds to $5$-minute interval. The cumulative DERA response corresponds to the output $y$ in \eqref{eq: system_output} when the system \eqref{eq: system} is driven by the policy learned via Algorithm \ref{alg: alg1} over $L\!=\!407$ episodes (blue line) and by the forecast-based MPC benchmark (dotted black line). We observe that the cumulative DERA responses track the net-load~demand.}
    \label{fig: tracking}
\end{figure}
We partition the data into $L$ episodes, each with a fixed time horizon $T\!=\!289$ time-steps corresponding to one day~of measurements, yielding $N_d\!=\!LT$. We consider $M\! =\! 5$ heterogeneous distributed energy resource aggregators --- air conditioners (ACs), electric water heaters (E-WHs), building HVACs (bldgs), refrigerators (RFGs), and electric vehicles (EVs) --- obeying the linear dynamics in \eqref{eq: system}, with the corresponding parameters summarized in Table \ref{table: parameters}. We assume the net load from the CAISO data evolves according to the linear Markov process in \eqref{eq: linApprox} with $d\!=\!3$ and $r\!=\!2$, where the measures $\mu_t$ are unknown. We choose $\phi(\cdot)$ as follows
\begin{align*}
 \phi(w_t)=
\begin{bmatrix}
f_1(w_t)/\left(f_1(w_t)+f_2(w_t)+f_3(w_t)\right)\\
f_2(w_t)/\left(f_1(w_t)+f_2(w_t)+f_3(w_t)\right)\\
f_3(w_t)/\left(f_1(w_t)+f_2(w_t)+f_3(w_t)\right)
\end{bmatrix},
\end{align*}
where
\begin{align*}
 f_1(w_t)=&1,\quad f_2(w_t)=\exp{\left(\frac{ \left(w_t-\nu_2\right)^{\T}\Sigma_2^{-1}\left(w_t-\nu_2\right)}{2r}\right)},\\
 f_3(w_t)=&\exp{\left(\frac{ \left(w_t-\nu_3\right)^{\T}\Sigma_3^{-1}\left(w_t-\nu_3\right)}{2r}\right)}.
\end{align*}
In order to choose the means $\nu_2\in \mathbb{R}^r$ and $\nu_3\in \mathbb{R}^r$ and the covariance matrices $\Sigma_2\in \mathbb{R}^{r\times r}$ and $\Sigma_3\in \mathbb{R}^{r\times r}$, we first re-write $S$ in \eqref{eq: net load data} as
\begin{align}\label{eq: net load data reshaped}
 S_r\!=\!
\begin{bmatrix}
 s_{-r+1} & s_{-r+2} & \cdots & s_{N_d-r+1}\\
 s_{-r+2} & s_{-r+3} & \cdots & s_{N_d-r+2}\\
 \vdots & \vdots & \ddots & \vdots \\
 s_{0} & s_{1} & \cdots & s_{N_d}
\end{bmatrix}
\!=\!
\begin{bmatrix}
 w_{0} & \cdots & w_{N_d}
\end{bmatrix}.
\end{align}
Then, we use the K-means algorithm over the data $w_t$ for $t\in \{0, \cdots , N_d\}$ using $d-1=2$ clusters to obtain the feature centers $\nu_2$ and $\nu_3$. For each cluster, we compute the sample covariance of the associated $w_t$ vectors, which yields the covariance matrices $\Sigma_2$ and $\Sigma_3$, respectively. We choose the weights of cost function in \eqref{eq: cost} as
\begin{align*}
 W_x= \blkdiag{\left(W_{x}^1,W_{x}^2,W_{x}^3,W_{x}^4,W_{x}^5\right)}
\quad
W_x^{i}=
\begin{bmatrix}
 \kappa_z^i & 0\\
 0 & \kappa_p^i
\end{bmatrix},
\end{align*}
where $\kappa_z^i$ and $\kappa_p^i$ are in Table \ref{table: parameters} for $i \in \{1,\cdots , 5\}$, $R=10^4 I_5$ and $\kappa=10^2$. We use Algorithm \ref{alg: alg1} to learn the weight parameters, $\theta$, that parametrizes the optimal policy in \eqref{eq: optimal policy}. We set $L=407$, $R_{\theta}=10^4$, and $\lambda=0.1$. For $\ell=1$, we set $x_0^1 \sim \mathcal{N}(0,I_{10})$, $\Lambda^1_t = \lambda I_{33}$, and $\theta^1_{t+1}\sim \mathcal{N}(0,I_{33})$, for $t\in\{0,\cdots, T-1\}$. During deployment, at each dispatch interval of $5$ minutes, the system operator broadcasts the net-load signal $s_t$, DER aggregators report their aggregate states, $x_{i,t}$, and the controller transmits ramp-rate commands, $u_{i,t}$. Since the controller in \eqref{eq: optimal policy} is in closed-form and requires only matrix-vector multiplications, the required latency is in the order of seconds. The framework avoids solving online optimization problems as in forecast-based MPC.

Fig. \ref{fig: tracking} shows the net-load demand (depicted by the dashed red line), the cumulative DERA response under the proposed forecast-free RL controller in Algorithm \ref{alg: alg1} (depicted by the solid blue line), and the cumulative DERA response under the forecast-based MPC benchmark (depicted by the dotted black line) over a five day period, where each time-step corresponds to $5$-minute interval. The cumulative DERA response corresponds to the output $y$ in \eqref{eq: system_output} when the system \eqref{eq: system} is driven by the policy learned via Algorithm \ref{alg: alg1} over $L\!=\!407$ episodes (blue line) and by the forecast-based MPC benchmark (dotted black line). We observe that both cumulative DERA responses track the net-load demand. While the forecast-based MPC benchmark can achieve lower tracking error due to access to short-term predictions, the proposed forecast-free RL controller achieves comparable tracking performance without requiring demand forecasts or online optimization. Moreover, the RL controller continuously adapts to changes in net-load statistics through online learning (Remark \ref{rmrk: data diversity}), whereas forecast-based approaches rely on fixed prediction models that must be retrained when data distributions shift.

\begin{figure}[!t]
  \centering
  \includegraphics[width=0.95\columnwidth,trim={0cm 0cm 0cm
    0cm},clip]{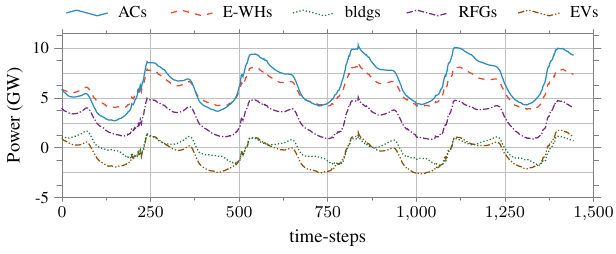}
  \caption{This figure shows the power trajectories of each DERA: ACs (depicted by the solid blue line), E-WHs (depicted by the dashed red line), bldgs (depicted by the dotted green line), RFGs (depicted by the dash-dotted purple line), and EVs (depicted by the dash-double-dotted brown line) over a five day period, where each time-step corresponds to $5$-minute interval. Each trajectory corresponds to the states $p_1$, $p_2$, $p_3$, $p_4$, and $ p_5$ of the system \eqref{eq: system}, respectively, when driven by the policy learned via Algorithm \ref{alg: alg1} over $L\!=\!407$~episodes.}
    \label{fig: DERA_input}
\end{figure}

\begin{figure}[!t]
  \centering
  \includegraphics[width=0.95\columnwidth,trim={0cm 0cm 0cm
    0cm},clip]{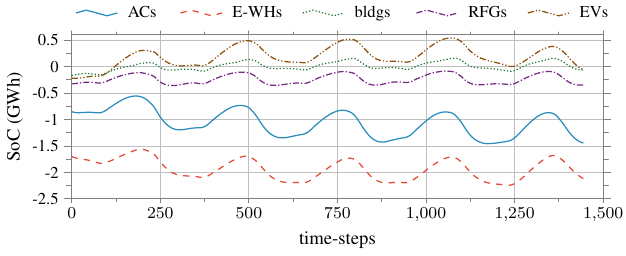}
  \caption{This figure shows the SoC trajectories of each DERA: ACs (depicted by the solid blue line), E-WHs (depicted by the dashed red line), bldgs (depicted by the dotted green line), RFGs (depicted by the dash-dotted purple line), and EVs (depicted by the dash-double-dotted brown line) over a five day period, where each time-step corresponds to $5$-minute interval. Each trajectory corresponds to the states $z_1$, $z_2$, $z_3$, $z_4$, and $ z_5$ of the system \eqref{eq: system}, respectively, when driven by the policy learned via Algorithm \ref{alg: alg1} over $L\!=\!407$~episodes.}
    \label{fig: DERA_SoC}
\end{figure}

\begin{figure}[!t]
  \centering
  \includegraphics[width=0.92\columnwidth,trim={0cm 0cm 0cm
    0cm},clip]{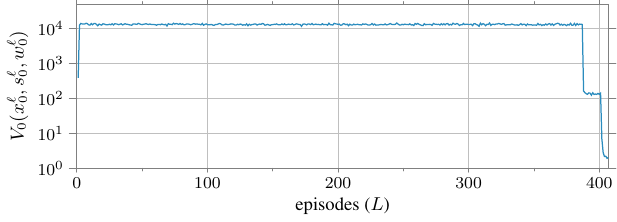}
  \caption{This figure shows the cumulative cost incurred by the policy learned from Algorithm \ref{alg: alg1} starting from the initial state $x_0^{\ell}$, $s_0^{\ell}$, and $w_0^{\ell}$ for episodes $1\leq \ell \leq 407$. For episodes $1\leq \ell \leq 387$, we add a random exploration term sampled from $\mathcal{N}(0,0.25 I_5)$ to the policy (given by line 13 of  Algorithm \ref{alg: alg1}), and for episodes $387 < \ell \leq 401$, we add a random exploration term sampled from $\mathcal{N}(0,0.0025 I_5)$. We observe that the cumulative cost decreases as the number of episodes increases after the exploration term is removed or reduced.}
    \label{fig: V0}
\end{figure}

Fig. \ref{fig: DERA_input} and Fig. \ref{fig: DERA_SoC} show the power and SoC trajectories of each DERA, respectively,  
over a five day period, where each time-step corresponds to a $5$-minute interval. These trajectories correspond to the state trajectories of the system \eqref{eq: system} when driven by the policy learned via Algorithm \ref{alg: alg1} over $L\!=\!407$ episodes. Fig. \ref{fig: V0} shows the cumulative cost incurred by the policy learned from Algorithm \ref{alg: alg1} starting from the initial state $x_0^{\ell}$, $s_0^{\ell}$, and $w_0^{\ell}$ for episodes $1\leq \ell \leq 407$. For $1\leq \ell \leq 387$, we add a random exploration term sampled from $\mathcal{N}(0,0.25 I_5)$ to the policy in Algorithm \ref{alg: alg1} line 13, and for $387< \ell \leq 401$, we add a random exploration term sampled from $\mathcal{N}(0,0.0025 I_5)$. We observe that the cumulative cost decreased as the number of episodes increases after the exploration term is removed.
\begin{remark}{\bf \emph{(Need for exploration)}}\label{rmrk: exploration}
In contrast to our earlier formulation in \cite{AAAM-OK-LS-SZ:25}, in this work, the initial states are not i.i.d., but instead inherit temporal dependence from the underlying Markov process carried over from the previous episode. This lack of episodic reset weakens the natural excitation present in the data. To compensate, we incorporate a mild exploration term to ensure convergence of the learned weights.
\end{remark}
\begin{remark}{\bf \emph{(Dataset diversity and robustness)}}\label{rmrk: data diversity}
The CAISO net-load dataset that we use spans May-September 2023, which captures seasonal transitions (e.g., solar ramps, heat waves, and peak demand events). Our control policy is based on a structured linear policy rather than a black-box function approximator (e.g, neural network), reducing risk of overfitting. Furthermore, explicit exploration (Remark \ref{rmrk: exploration}) and online episodic policy update ensure sufficient excitation and robustness to temporal correlations.
\end{remark}

\begin{table}[t]
\vspace{-0.75 em}
	\centering
	\begin{tabular}{|| l c c c c c r ||}
		Par. & Unit & DER1 & DER2 & DER3 & DER4 & DER5 \\
		Type & --- & ACs & E-WHs & bldgs & RFGs & EVs \\
		N & million & 10 & 10 & 1 & 10 & 1   \\
		$\alpha_i$ & ---  & 0.98 & 0.99 & 0.97 & 0.96 & 0.99  \\
		$\kappa^i_z$ & --- & 1e-3 & 2e-3 & 5e-3 & 5e-3 & 2e-3 \\
		$\kappa^i_p$ & --- & 1e-1 & 2e-1 & 5e-1 & 5e-1 & 2e-1 \\
            $\beta_i$ & hr & 1/300 & 1/300 & 1/300 & 1/300 & 1/300
	\end{tabular}
	\bigskip
\caption{Parameters for each class of DER, corresponding to the discrete-time linear dynamical model in \eqref{eq: system}.}\label{table: parameters}
\end{table} 
\section{Conclusion}\label{sec: conclusion}
In this paper, we develop a forecast-free RL framework for enabling heterogeneous distributed energy resource aggregators, DERAs, to track net-load demand without relying on short-term predictions. By coupling deterministic linear DERA dynamics with a feature-based linear Markov model for the net load, we derived closed-form expression for the optimal policy. This expression enabled a least-squares value iteration scheme that learns directly from operational data while preserving the interpretability and structure of classical LQR-type controllers. Numerical experiments using real CAISO net-demand data demonstrated that the learned controller achieves high tracking accuracy, smooth actuation, and stable regulation across DER classes. Our results show that reliable real-time DERA allocation can be achieved without explicit load forecasting, highlighting the value of hybrid model-based and data-driven control frameworks for modern grid-balancing applications.

\end{document}